# *Synthesis of thin-film black phosphorus on a flexible substrate*

Xuesong Li[1†], Bingchen Deng[1†], Xiaomu Wang[1], Sizhe Chen[2], Michelle Vaisman[1],

Shun-ichiro Karato[3], Grace Pan[4], Minjoo Larry Lee[1], Judy Cha[4], Han Wang[2], and

Fengnian Xia[1*]

[1]*Department of Electrical Engineering, Yale University, New Haven, Connecticut 06511*
[2]*Ming Hsieh Department of Electrical Engineering, University of Southern California, Los Angeles, CA 90089*
[3]*Department of Geology and Geophysics, Yale University, New Haven, Connecticut 06511*
[4]*Department of Mechanical Engineering and Materials Science, Yale University, New Haven, Connecticut 06511*

[†]These authors contributed equally to this work.

*[*fengnian.xia@yale.edu](mailto:fengnian.xia@yale.edu)

**Abstract**

We report a scalable approach to synthesize a large-area (up to 4 mm) thin black phosphorus (BP) film on a flexible substrate. We first deposited a red phosphorus (RP) thin-film on a flexible polyester substrate, followed by its conversion to BP in a high-pressure multi-anvil cell at room temperature. Raman spectroscopy and transmission electron microscopy measurements confirmed the formation of a nano-crystalline BP thin-film with a thickness of around 40 nm. Optical characterization indicates a bandgap of around 0.28 eV in the converted BP, similar to the bandgap measured in exfoliated thin-films. Thin-film BP transistors exhibit a field-effect mobility of around 0.5 $cm^2$/Vs, which can probably be further enhanced by the optimization of the conversion process at elevated temperatures. Our work opens the avenue for the future demonstration of large-scale, high quality thin-film black phosphorus.

Orthorhombic black phosphorus (BP) has recently attracted significant attention in the layered material community due to its high mobility, widely tunable electronic bandgap by layer thickness from 0.3 eV in bulk to ~ 2 eV in a monolayer[1-5], and its highly anisotropic in-plane electronic and optical properties[6]. These desirable properties make BP a potential candidate for various nanoelectronic and nanophotonic applications[7-11]. In addition to the orthorhombic phase, which is stable under ambient conditions, BP can also be in the rhombohedral or simple cubic phase[12]. Under high pressure, BP reversibly transits from orthorhombic into rhombohedral at about 5 GPa at room temperature and then simple cubic at about 10 GPa, and the latter phase transition process is almost independent of temperature[13]. In 1914 Bridgman first reported that white phosphorus (WP) can be converted into BP under a hydrostatic pressure of 1.2 GPa at 200 °C within 5 to 30 min[14]. He later prepared BP at room temperature from WP at about 3.4 GPa[15]. Conversion from red phosphorus (RP) to BP at 8.5 GPa was also demonstrated by him at room temperature[16]. Rissi *et al*. showed that at this pressure RP first converts into rhombohedral BP and finally into orthorhombic BP after decompressing[17]. The pressure for RP to BP conversion can be as low as 3.8-4.7 GPa with shear stress[18, 19]. Sorgato *et al*. investigated the conversion from RP to BP as a function of temperature at hydrostatic pressures up to 6.5 GPa and temperatures up to 560 °C[20]. By fitting their data, they derived a linear relationship between temperature and hydrostatic pressure under which the conversion from RP to BP occurs in about 10 min, $T = T_0 + \alpha P$, where $T_0$ = 560 °C and $\alpha$ = -63.5 °C /GPa. The above mentioned phase transition conditions for synthesis of BP under different pressures are summarized in Fig. 1. Extrapolating Sorgato *et al*.'s experimental results to room temperature (as shown by the red line in Fig. 1) leads to a phase transition pressure of around 8 GPa, in good agreement with Bridgeman's results obtained from hydrostatic pressurization.

Furthermore, by melting BP at a temperature of 900 ºC and under a hydrostatic pressure of 1 GPa, BP single crystal larger than 5 × 5 × 10 mm$^3$ was achieved, as reported by Endo *et al.* in 1982[21].

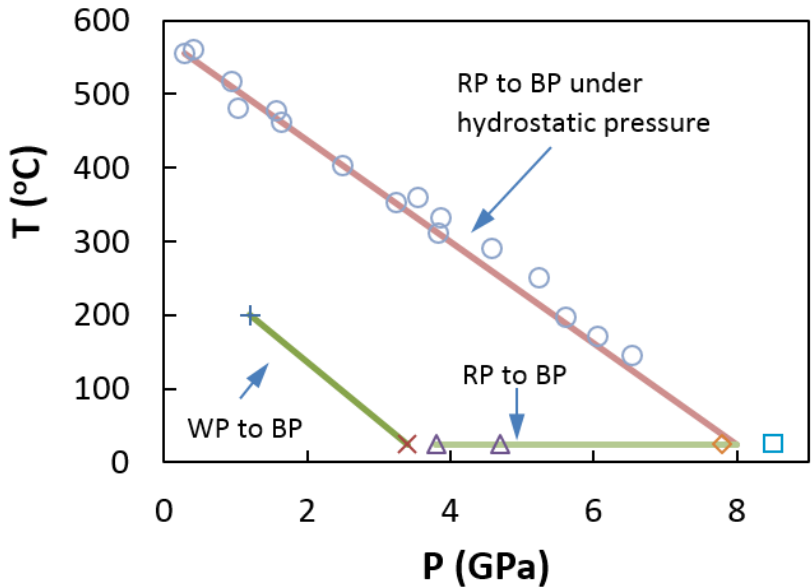

**Figure 1**. Criteria for the conversion of BP from WP and RP at various pressures and temperatures summarized from the literature: +[14], ×[15], △[18,19], □[16], ◇[17], ○[20].

Alternative BP synthesis techniques without using high pressure were also developed, such as the technique involving mercury as a catalyst developed by Krebs *et al.* in 1955[22], the bismuth-flux based method by Brown *et al.* in 1965[23], and the mineralizer-assisted short-way transport reaction method by Lange *et al.* in 2007[24] and Nilges *et al.* in 2008[25], and further optimized by Kopf *et al.* in 2014[26]. However, it should be noted that all above mentioned methods are for the synthesis of bulk BP while few-layer and thin film BP so far can only be prepared by the micro-mechanical cleavage method, which is not scalable for large-area production.

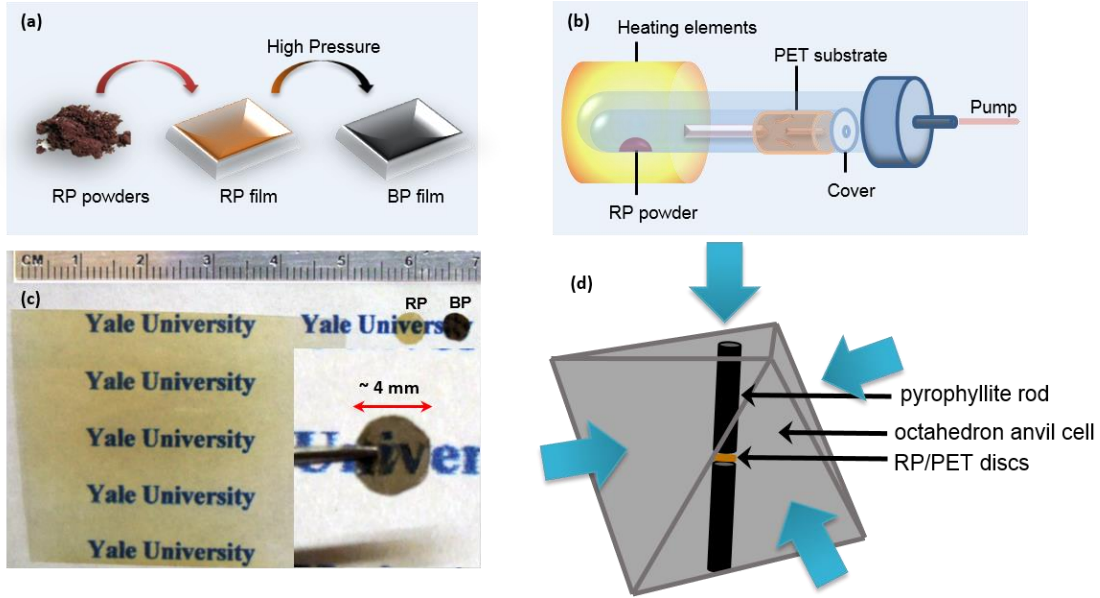

**Figure 2.** (a) Strategy for the synthesis of thin BP films. (b) Schematic apparatus for the deposition of RP film. (c) Photos of thin RP film on PET substrate (left), RP/PET disc for pressurization (middle) and BP/PET disc after pressurization. The inset shows the transparency of the BP/PET film. (d) Schematic of the high pressure anvil cell for conversion. The arrows indicate the directions along which the pressure is applied in conversion process.

Here we report a method towards large-area synthesis of a BP thin film on flexible substrates as illustrated in Fig. 2a: a thin RP film is first deposited onto the substrate, which is subsequently converted into a BP film by pressurization in an anvil cell. The conversion process was carried out at room temperature, at which a pressure beyond 8 GPa is required for complete conversion from RP to BP, as shown in Fig. 1. RP film can be deposited by sputtering[27] or thermal deposition[28]. Here we prepared the RP film by thermal deposition in a self-designed apparatus, which is schematically shown in Fig. 2b. About 5 milligrams of RP powders (Sigma-Aldrich, purity ⩾ 99.99%) were placed inside a quartz tube with one open end. A piece of 40 mm × 50 mm × 75 μm polyethylene terephthalate (PET) film was wrapped on the inside wall of the tube close to the open end. The open end of the tube was then covered by a thin polyester film with a ~0.5-mm diameter pin hole at the center. The PET film was chosen due to its intrinsic flexibility and inertness with phosphorus. The cover was used to restrict the phosphorus vapor flow so that the deposition of phosphorus onto the substrate can be efficient. The entire assembly was then loaded into a 50-mm diameter quartz tube chamber, with the powders at the center of the heating zone and the substrate at the room temperature region outside of the heating zone. The reaction chamber was then evacuated by a mechanical pump to a pressure under 10 mTorr. The RP powders were then heated up to 400 ºC in about 10 minutes and held at this temperature. The RP evaporates and then condenses onto the substrate to form a thin film, whose thickness is controlled by the holding time at 400 ºC.

The transparent PET film before deposition turns to red-brown after phosphorus deposition in the chamber for 40 minutes, as shown in Fig. 2c left. The red-brown color indicates that the deposited film is probably major RP but not WP. The film is stable in air, which further indirectly proves it is not WP since WP combusts spontaneously in air. The film is continuous and uniform over most region of the substrate except for the edges. Longer deposition time results in thicker RP film evidenced by the decrease of the transparency. The RP/PET film was then cut into 4.4-mm diameter discs (on top right of Fig. 2c), sandwiched by two pyrophyllite rods with the same diameter and then loaded into a chromium doped magnesium oxide (MgO) octahedron pressure medium for pressurization (schematically shown in Fig. 2d) in a 1000 ton Kawai-type multi-anvil apparatus. Multi-anvil apparatus is commonly used to produce high pressure in a small volume to simulate the hydrostatic pressures which is ubiquitous within planets[29]. The pressure in the multi-anvil cell was increased to 10 GPa in 4 hours, then held for 6 hours, and finally released to ambient pressure in 9 hours. Multiple RP discs (up to 10) were stacked together and loaded into the anvil chamber in a single run. After high-pressure treatment, the diameter of the discs shrinks slightly (~10%) and the color becomes semi-transparent black (on top right Fig. 2c), which can be seen more clearly from a close-up view of the converted BP disc (inset of Fig. 2c). The converted BP samples were then stored in a nitrogen box, and no obvious degradation was observed three months after conversion.

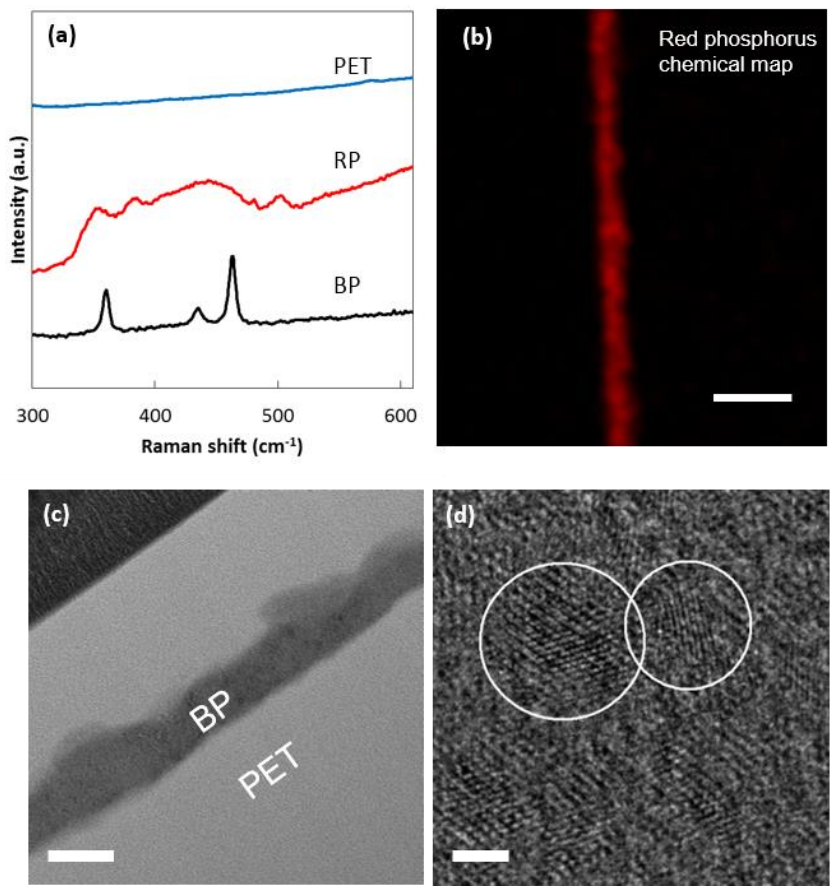

**Figure 3.** (a) Raman spectra of the bare PET substrate, RP on PET, and BP on PET, respectively. (b) The EDX chemical map of P from the as-grown RP thin film. (c) and (d) The low and high resolution transmission electron microscopy images of the BP cross-section. Scale bars for (b), (c), and (d) are 200 nm, 50 nm, and 5 nm, respectively.

We further performed various spectroscopy measurements to characterize the RP and BP thin films. Fig. 3a denotes the Raman spectra of the substrate, the as-deposited RP film and the converted BP thin film. The excitation wavelength is 532 nm. A 50 x objective was utilized to focus the beam to a spot with a diameter of around 1.5 μm. There are no prominent features from the PET substrate in the investigated region (300 – 610 cm$^{-1}$). The Raman spectrum from the as-deposited P film shows typical RP characteristic peaks at ~350, 380, 440, and 500 cm$^{-1}$, respectively, indicating that the film is indeed comprised primarily of RP[27], and not WP[30]. The relatively low intensity of the peak at 350 cm$^{-1}$ indicates that the RP film is disordered[31]. Although it is known that below 800 °C phosphorus vapor exists in the form of $P_4$ molecules[32], which condenses to form a mixture of WP and RP[33], in this work we attribute the formation of RP film from the conversion of WP since the film is very thin and WP can be easily converted into RP after exposure to ambient light in the quartz growth chamber. On the other hand, the Raman spectrum from the film after pressurization shows typical orthorhombic BP characteristic peaks at ~360, 435, and 463 $cm^{-1}$, respectively[17, 34], clearly indicating the RP to BP conversion after pressurization. Moreover, tens of Raman spectra taken randomly over the whole sample all showed similar characteristics, indicating a uniform conversion of RP to BP. A chemical map of the as-grown RP film was taken by energy dispersive X-ray spectroscopy in a scanning transmission electron microscope (STEM), which shows a uniform coverage of a ~ 40 nm thick RP film on the flexible substrate (Fig. 3b). Cross-sectional TEM samples of the RP thin films were prepared by mechanical wedge polishing, followed by ion milling as a final step. The crystalline nature of the converted BP film was confirmed by TEM. Fig. 3c shows a low magnification TEM image of the cross section of the BP/PET film. The thickness of the BP film is about 40 nm, similar to the thickness of RP before conversion. High-resolution TEM image indicates that the BP grain size is in the order of 10 nm (Fig. 3d). For BP TEM samples, focused ion beam was used due to delamination issues during mechanical polishing of the samples.

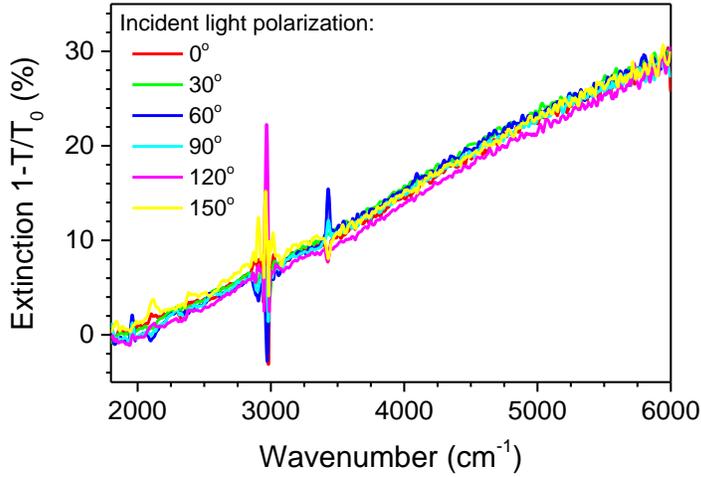

**Figure 4.** Polarization resolved optical extinction spectra of converted BP thin-film.

The optical property of the BP film is characterized using Fourier-transform infrared spectroscopy (FTIR). We plotted extinction in transmission, $1-T/T_0$, to characterize the infrared response of the BP thin-film under different incident light polarization (Fig. 4), where $T$ is the transmission of the light through the BP film and the PET substrate, while $T_0$ is the transmission through the PET only. The results qualitatively agree with the spectra obtained from exfoliated BP[6], although no in-plane anisotropy is observed due to the nanoscale grains with random crystalline orientations. The BP phase is further confirmed by the rise of absorption at ∼ 2200 cm$^{-1}$, indicating a bandgap of around 0.28 eV, close to the bandgap of 0.3 eV observed in exfoliated flakes. The kinks near 2970 cm$^{-1}$ and 3430 cm$^{-1}$ result from vibrations of hydroxyl (-OH) and methyl (-CH$_3$) groups in the PET substrate, respectively.[35]

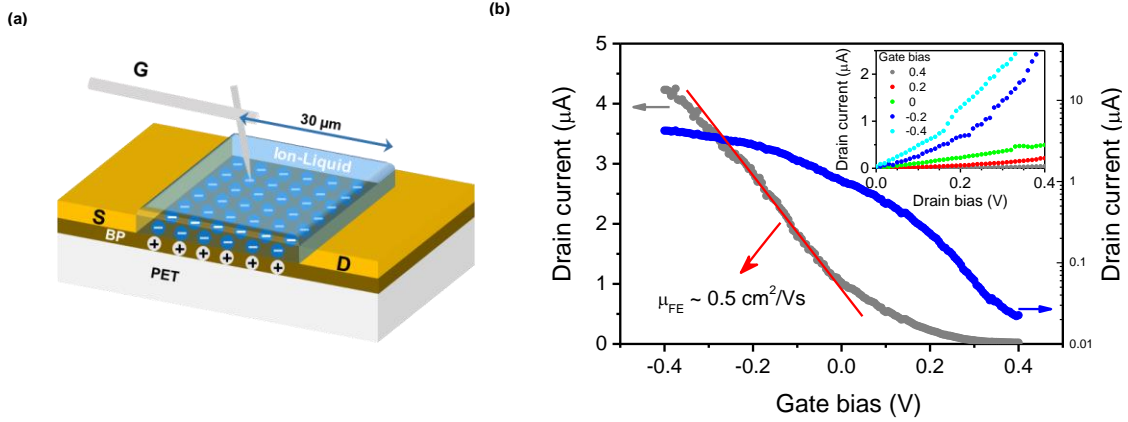

**Figure 5.** (a) Schematic of the BP FET device. (b) Transfer characteristics of a transistor made from converted BP thin-film in linear (left axis) and logarithmic (right axis) scales. Inset: Output characteristics of the same transistor.

We also fabricated BP field-effect transistors (FET) directly on the flexible PET substrate to characterize its transport properties. Source and drain electrodes were deposited using a shadow mask. The transistor channel length and width are 30 and 70 μm, respectively. After metallization, an electrochemical gate was formed using ionic liquid (0.1 mol/L LiClO$_4$ aqueous solution).[36, 37] A gold wire was attached to the ionic liquid as a gate. A schematic of the FET device is shown in Fig. 5a. The transfer characteristics (at 0.25 V source-drain bias) are plotted in Fig. 5b, indicating that the converted BP is p-type. The transistor on/off ratio is around 200. The carrier field-effect mobility in the linear region can be estimated using

$$\mu = \frac{L}{W}\frac{1}{C_i}\frac{\partial g_{ds}}{\partial V_g}$$

where $L$ and $W$ are length and width of the transistor channel, respectively, $g_{ds}$ is the conductance measured between the source and drain, $C_i$ is the capacitance of the ionic liquid gate, and $V_g$ is the gate bias. The $C_i$ is estimated to be around 15 $\mu F/cm^2$ in our devices.[37] The hole mobility is calculated to be 0.5 cm$^2$/Vs using the parameters above. The reported mobility here is lower than that in exfoliated flakes, due to the small grain size of the converted BP. Moreover, the BP transistors reported here show typical p-type behavior, which can be due to factors such as the initial BP doping after conversion and environmental doping effect. In fact, many BP transistors made from exfoliated flakes also exhibit p-type behavior[6]. Finally the current fluctuation in the output characteristics of BP transistors as shown in the inset of Fig. 5b is probably due to absorption and desorption of air molecules, a phenomenon observed previously in carbon nanotube transistors[38, 39].

Although in this work the individual converted BP film is only several millimeters in diameter, our approach is scalable for the production of large area BP films. Firstly we were able to produce up to 10 thin films in one run, and in principle, more RP films can be converted to BP as long as the RP stack can fit in the anvil cell. Secondly, the RP film can be deposited on much larger substrates by scaling up the size of the chamber. Although generating high pressure beyond 8 GPa within a large area is challenging, the flexible substrate can be wrapped or folded to fit in the centimeter scale anvil cell. It should be noted that in our method, RP films can be deposited on any substrate that is inert to phosphorus. For example, RP films can be deposited on a silicon wafer, and then the phase transition can be accomplished by immersing the substrate in liquid and applying hydrostatic pressure. Although our converted BP thin film exhibits a relatively low mobility compared to the exfoliated samples, raising the temperature during the conversion process is expected to enhance the BP grain size significantly due to the increased mobility of phosphorus atoms, which will lead to much enhanced carrier mobility. Moreover, as shown in Fig. 1, higher conversion temperatures can also reduce the pressure required for the phase transition as long as a thermostable substrate is chosen, and a larger anvil chamber can then be utilized to accommodate larger substrates. The estimated deposition rate of RP can be as lower as around 1 nm per minute and the roughness of the film is usually in the order of 2 to 3 nm. As a result, we expect that uniform RP films as thin as ~ 5 nm can be achieved using this approach. If an inert substrate which can sustain a high temperature beyond the melting point of BP (> 900 ℃) is used during the conversion process, wafer-scale single crystalline thin-film black phosphorus may be realized. Finally for device applications, it is highly desirable to directly grow thin film material on flexible substrates. Growth of thin film BP on polyester films represents an important step toward the realization of high performance, bendable electronics and photonics.

In summary, we demonstrate a scalable approach for the synthesis of large-area (up to 4 mm) black phosphorus thin films with a thickness of around 40 nm. Red phosphorus thin films were firstly deposited on a flexible substrate followed by conversion to black phosphorus thin films in a high-pressure multi-anvil cell at room temperature. Spectroscopic characterizations reveal that the converted material is nano-crystalline black phosphorus. Optical characterization indicates a bandgap of around 0.28 eV, close to the bandgap of exfoliated BP thin-films. Thin-film BP transistors exhibit a field-effect mobility of around 0.5 cm$^2$/Vs, which can probably be enhanced by future optimization of BP conversion process. Our work demonstrates feasibility of future realization of wafer-scale, high quality thin-film black phosphorus.

**Acknowledgements:**
F. X. acknowledges support from the Office of Naval Research (N00014-14-1-0565) and the National Science Foundation (CRISP NSF MRSEC DMR-1119826). We thank Kanani Lee, George Amulele, and Jennifer Girard of Department of Geology and Geophysics at Yale University for technical assistance.